\documentclass[10pt, aps, prd, amsmath, floats, floatfix, twocolumn, notitlepage,
superscriptaddress, nofootinbib, showpacs]{revtex4-2}
\allowdisplaybreaks
\usepackage{float}
\usepackage[utf8]{inputenc}
\usepackage[T1]{fontenc}
\usepackage{comment}
\usepackage{bm}
\usepackage[normalem]{ulem}\usepackage{mathtools,amsmath,amssymb,amsfonts,mathrsfs,eucal,graphicx,tensor,csquotes,accents,commath,chngcntr,siunitx}
\usepackage[dvipsnames]{xcolor}
\usepackage[unicode]{hyperref}
\usepackage{orcidlink}
\hypersetup{colorlinks=true, citecolor=MidnightBlue,
            linkcolor=MidnightBlue, urlcolor=MidnightBlue, linktocpage=true}
\usepackage[normalem]{ulem}

\DeclareMathAlphabet{\mathpzc}{OT1}{pzc}{m}{it}

\definecolor{darkgreen}{rgb}{0.0, 0.6, 0.0}

\usepackage{tikz}
\usepackage{hyperref}
\usepackage{graphicx} 
\usetikzlibrary{shapes.geometric, arrows.meta, positioning, calc}

\newcommand{\cH}{\mathcal H}
\newcommand{\cS}{\mathcal S}
\newcommand{\Lie}{\mathcal L}

\begin{document}

\title{
How Universal Is the Black Hole Zeroth Law?}

\author{Rajes Ghosh~\orcidlink{https://orcid.org/0000-0002-1264-938X}}
\email[]{rghosh13@jh.edu}
\affiliation{William H. Miller III Department of Physics \& Astronomy, Johns Hopkins University\\
3400 North Charles Street, Baltimore, MD 21218, USA}

\author{Sudipta Sarkar~\orcidlink{https://orcid.org/0000-0002-4077-1358}}
\email[]{sudiptas@iitgn.ac.in}
\affiliation{Department of Physics, Indian Institute of Technology Gandhinagar,
Palaj, Gandhinagar 382355, India}

\date{\today}

\begin{abstract}
The black hole zeroth law asserts the thermal equilibrium of a stationary Killing horizon, yet its status beyond general relativity remains limited. We demonstrate that its validity is not governed by a universal mechanism: distinct gravitational theories enforce constancy of the surface gravity through different horizon-projections of their field equations. In general relativity, $f(R)$-gravity, and Lovelock theories, the zeroth law is known to follow from the mixed (null-transverse) horizon equations. By contrast, we show that in non-Lovelock theories the zeroth law is enforced by qualitatively different mechanisms. In generalized quadratic gravity, for instance, the null-null component of the field equations dictates its validity. Exploiting this novel diagnostic, we obtain the first exact proof of the zeroth law in the broader landscape of higher-curvature gravity, without relying on a perturbative expansion around general relativity. Our result establishes the zeroth law as a sharp probe of the underlying thermodynamic structure of gravity, and points toward a classification of modified theories according to the mechanism responsible for its validity.
\end{abstract}

\maketitle

\section{Introduction}
The laws of black hole (BH) mechanics provide a fundamental link among gravitation, thermodynamics, and quantum field theory~\cite{Wald:1999vt, Carlip:2014pma, Wall:2018ydq, Sarkar:2019xfd}. Central to this connection is the zeroth law: the surface gravity $\kappa$ is constant on a stationary BH horizon. In general relativity (GR), this follows from Killing-horizon geometry, Einstein's field equations, and, in the presence of matter, suitable energy conditions~\cite{BardeenCarterHawking}. Through Hawking's relation between temperature and surface gravity~\cite{Hawking:1974rv,Hawking:1975vcx}, the zeroth law is also the condition that allows a stationary BH to represent a thermodynamic system in thermal equilibrium.

It is therefore natural to ask how universal this equilibrium principle is beyond GR. This question is particularly relevant in the era of gravitational-wave astronomy, where deviations from GR are being actively constrained~\cite{Yunes:2009ke, Agathos:2013upa, Berti:2015itd, LIGOScientific:2016lio, Isi:2019aib, LIGOScientific:2021sio, LIGOScientific:2026qni, LIGOScientific:2026oim}. Higher-curvature interactions provide a natural arena for such tests, arising both in effective descriptions of gravity and in candidate ultraviolet completions. The observational tests of the second law of BH mechanics~\cite{Isi:2020tac,LIGOScientific:2025rid} and the recent developments on the thermodynamics-based understanding of binary merger end states~\cite{Rincon-Ramirez:2026tbo} further motivate the question of whether the zeroth law also survives beyond Einstein gravity.

Considerable progress has been made in extending the zeroth law to modified theories~\cite{Jacobson:1995uq, GhoshSarkar, Dey:2021rke, Sang:2021rla, Fang:2022nfa, Bhattacharyya:2022nqa, Davies:2024fut, Sang:2024tjd}. However, most general results are perturbative in the higher-curvature couplings and therefore apply to solutions continuously connected to that of GR. Such arguments cannot capture genuinely non-Einstein branches, which occur in several modified theories. Quadratic gravity (QG), for example, admits exact non-Einstein BH solutions~\cite{Lu:2015cqa, Lu:2015psa, Pravda:2016fue, Kokkotas:2017zwt, Podolsky:2019gro, Held:2022abx}, encouraging a non-perturbative and branch-independent analysis of the zeroth law~\cite{note, Ghosh:2024tlk}.

The problem is nontrivial already in Lovelock gravity. Although Lovelock theories retain second-order field equations~\cite{Padmanabhan:2013xyr}, non-perturbative zeroth-law arguments encounter an algebraic obstruction controlled by the intrinsic geometry of the horizon cross section~\cite{SarkarBhattacharya,GhoshSarkar}. Generic higher-derivative theories admit perturbative proofs order-by-order in the couplings~\cite{Bhattacharyya:2022nqa}, but no comparable branch-independent mechanism is known in general. This leaves open a more structural question: \emph{which part of the gravitational field equations actually enforces horizon equilibrium?}

In this work, we show that the answer is theory dependent. We prove an exact non-perturbative zeroth law for generalized quadratic gravity (GQG)~\cite{Edelstein:2024jzu}, without expanding in the higher-curvature couplings or assuming a GR-connected branch. More generally, we show that different theories enforce constancy of $\kappa$ through different horizon projections of their field equations $E_{ab}$. Our main diagnostic is the null-null component $E_{ab}\xi^a\xi^b\equiv E_{\xi\xi}$ evaluated on a stationary Killing horizon $\mathcal H$ generated by $\xi^a$. In GR, $f(R)$ gravity, and Lovelock theories, this projection is insensitive to transverse variations of $\kappa$, and the zeroth law is instead controlled by the mixed (null-transverse) component $E_{ab}\xi^ae_A^b\equiv E_{\xi A}$. 

\section{Horizon geometry and identities}
We consider a smooth stationary Killing horizon $\cH$ with basis $\{\xi^a,N^a,e_A^a\}$, where $\xi^a$ is the null generator, $N^a$ is an auxiliary null vector satisfying $\xi\!\cdot\!N=-1$, and $e_A^a$ span a $(D-2)$-dimensional spatial cross section $\cS$. The induced metric is $\sigma_{AB}$, with $e_A^ae_B^b\sigma^{AB}=g^{ab}+2\xi^{(a}N^{b)}$. Capital Latin indices label directions tangent to $\cS$, while lower-case Latin indices denote spacetime directions.

The surface gravity is defined by $\xi^a\nabla_a\xi^b=\kappa\, \xi^b$. Stationarity implies $\Lie_\xi\kappa=0$, so proving the zeroth law reduces to showing the vanishing of the transverse variation $D_A\kappa$ on $\cH$. The vanishing expansion and shear of the Killing generators imply the standard horizon identities 
\begin{equation}
    R_{\xi\xi}=R_{\xi A\xi B}=R_{\xi ABC}=0,
    \quad R_{\xi A}=-D_A\kappa .
\label{EM:horizon}
\end{equation}
Thus, transverse variations of the surface gravity are directly encoded in the mixed (null-transverse) Ricci projection.

\section{Non-perturbative proof of the zeroth law}

For a general theory with field equations $E_{ab}=8\pi G\,T_{ab}$, our main diagnostic is the null-null projection $E_{\xi\xi}$ on $\cH$. Also, in the presence of matter, we use the no-horizon flux condition, namely $T_{\xi\xi}=0$ (as matter is invariant under the Killing flow). In theories where the standard GR/Lovelock strategy applies, one must also analyze the mixed projection $E_{\xi A}$, which vanishes under the dominant energy condition $T_{\xi A}=0$.

We start with the null-null projection $E_{\xi\xi}$. In GR, $f(R)$-gravity, and Lovelock theories, this projection vanishes identically off-shell on a stationary Killing horizon and therefore carries no information about $D_A\kappa$. In contrast, GQG realizes a qualitatively different possibility, whose Lagrangian is
\begin{equation*}
\mathcal{L}_{\textrm{GQG}}=R-2\,\Lambda+\alpha\, R^2+\beta\, R_{ab}R^{ab}
+\gamma\, \mathcal{L}_{\textrm{GB}},
\end{equation*}
where $\mathcal{L}_{\textrm{GB}}=R^2-4\,R_{ab}R^{ab}+R_{abcd}R^{abcd}$. We keep all couplings arbitrary, make no small-coupling expansion, and do not assume a GR-connected branch. On the Killing horizon, the $R^2$ and Gauss-Bonnet contributions to $E_{\xi\xi}$ vanish identically due to Eq.~\eqref{EM:horizon}~\cite{SarkarBhattacharya}, leaving an elegant and simple result,
\begin{equation}
E_{\xi\xi}=-\beta D^2(\kappa^2).
\label{eq:intro}
\end{equation}
For more details, check \ref{appA}. Using the GQG field equation and no-horizon-flux condition, it provides an intrinsic elliptic constraint on $\cH$,
\begin{equation*}
-\beta\, D^2(\kappa^2)=8\pi G\,T_{\xi\xi}=0.
\end{equation*}
On a compact horizon cross section without boundary, $\kappa^2$ is therefore a harmonic function when $\beta \neq 0$. Then, a straightforward application of the Hopf maximum principle~\cite{Jost}, implies $D_A \kappa=0$ for a connected non-extremal horizon, leading to the zeroth law. One can arrive at the same conclusion by multiplying $D^2(\kappa^2)=0$ by $\kappa^2$ and integrating by parts:
\begin{equation}
\int_{\cS}\sqrt \sigma\,
D_A(\kappa^2)\, D^A(\kappa^2)=0.
\end{equation}
This implies $D_A(\kappa)=0$ on $\cH$. Thus, unlike Lovelock theories, GQG admits an exact non-perturbative zeroth-law proof from the null-null projection of the field equations! 

Once constancy of $\kappa$ is established for $\beta\neq0$, the R\'acz-Wald extension gives a regular bifurcate Killing horizon~\cite{Racz:1995nh}, on which $\xi^a$ vanishes, and as a result, $E_{\xi A}$ vanishes too. Then, $E_{\xi A}=0$ everywhere on the horizon by Lie transport. Note, the exceptional sector $\beta=0$, which includes GR, $f(R)$-gravity, and Gauss-Bonnet theory, instead requires the mixed projection $E_{\xi A}$ for proving the zeroth law either non-perturbatively (GR and $f(R)$-gravity) or perturbatively (Lovelock theories)~\cite{Jacobson:1995uq, GhoshSarkar}. 

These observations suggest a novel classification of gravitational theories according to the horizon projections responsible for enforcing the zeroth law.
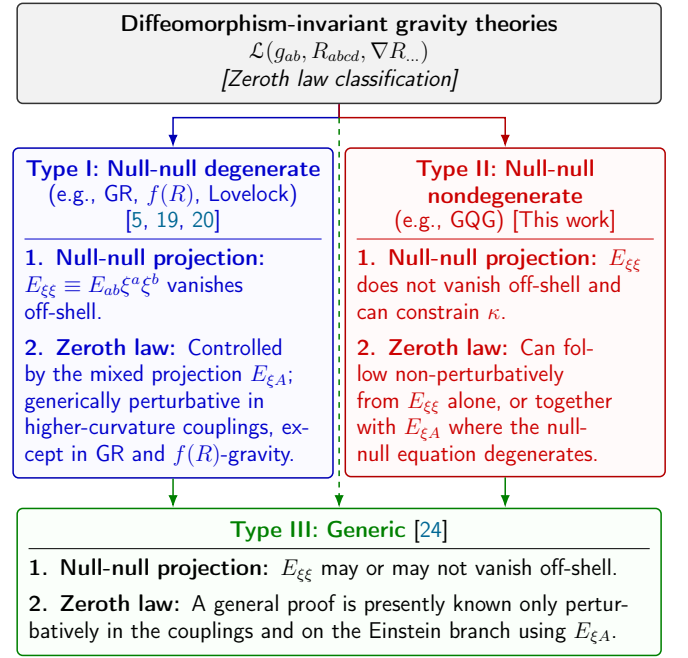
\begin{figure}[htbp]
\centering
\resizebox{\linewidth}{!}{%
\begin{tikzpicture}[
    >=Latex,
    font=\sffamily\large,
    titlebox/.style={
        rectangle,
        draw=black!80,
        fill=gray!10,
        thick,
        rounded corners=3pt,
        inner sep=6pt,
        align=center,
        text width=11.5cm
    },
    classIbox/.style={
        rectangle,
        draw=blue!70!black,
        fill=blue!2,
        thick,
        rounded corners=3pt,
        inner sep=6pt,
        text width=5.5cm,
        align=left
    },
    classIIbox/.style={
        rectangle,
        draw=red!70!black,
        fill=red!2,
        thick,
        rounded corners=3pt,
        inner sep=6pt,
        text width=5.5cm,
        align=left
    },
    classIIIbox/.style={
        rectangle,
        draw=green!50!black,
        fill=green!2,
        thick,
        rounded corners=3pt,
        inner sep=6pt,
        text width=11.5cm,
        align=left
    }
]

\node[titlebox] (root) {\centering
    \textbf{Diffeomorphism-invariant gravity theories}
    $\mathcal{L}(g_{ab},R_{abcd},\nabla R_{\dots})$\\[1pt]
    \itshape [Zeroth law classification]
};

\node[classIbox, below left=0.8cm and 0.1cm of root.south] (classI) {
    \centering\textbf{\color{blue!80!black} Type I: Null-null degenerate}\\[-1pt]
    \centering\color{blue!80!black}(e.g., GR, $f(R)$, Lovelock)
    \cite{BardeenCarterHawking, Jacobson:1995uq, GhoshSarkar}\\[3pt]
    \hrule height 0.4pt\vspace{5pt}
    \raggedright

    \textbf{1. Null-null projection:}
    $E_{\xi\xi}\equiv E_{ab}\xi^a\xi^b$ vanishes off-shell. \hfill\\[5pt]

    \textbf{2. Zeroth law:}
    Controlled by the mixed projection $E_{\xi A}$; generically perturbative in higher-curvature couplings, except in GR and $f(R)$-gravity. \hfill
};

\node[classIIbox, below right=0.8cm and 0.1cm of root.south] (classII) {
    \centering\textbf{\color{red!80!black} Type II: Null-null nondegenerate}\\[-1pt]
    \centering\color{red!80!black}(e.g., GQG) [This work]\\[3pt]
    \hrule height 0.4pt\vspace{5pt}
    \raggedright

    \textbf{1. Null-null projection:}
    $E_{\xi\xi}$ does not vanish off-shell and can constrain $\kappa$. \hfill\\[5pt]

    \textbf{2. Zeroth law:}
    Can follow non-perturbatively from $E_{\xi\xi}$ alone, or together with $E_{\xi A}$ where the null-null equation degenerates. \hfill
};

\node[classIIIbox, below=0.55cm of $(classI.south)!0.5!(classII.south)$] (classIII) {
    \centering\textbf{\color{green!50!black} Type III: Generic}~\cite{Bhattacharyya:2022nqa}\\[3pt]
    \hrule height 0.4pt\vspace{5pt}
    \raggedright

    \textbf{1. Null-null projection:}
    $E_{\xi \xi}$ may or may not vanish off-shell. \hfill\\[5pt]

    \textbf{2. Zeroth law:}
    A general proof is presently known only perturbatively in the couplings and on the Einstein branch using $E_{\xi A}$. \hfill
};

\coordinate (junction) at ($(root.south) + (0,-0.25)$);

\draw[->, thick, draw=blue!70!black] (root.south) -- (junction) -| (classI.north);
\draw[->, thick, draw=red!70!black] (root.south) -- (junction) -| (classII.north);
\draw[->, thick, draw=green!50!black] (classI.south) -- (classI.south |- classIII.north);
\draw[->, thick, draw=green!50!black] (classII.south) -- (classII.south |- classIII.north);
\draw[->, thick, draw=green!50!black, dashed] (junction) -- (classIII.north);

\end{tikzpicture}%
}
\caption{Classification of gravity theories based on the origin of the zeroth law.}
\label{fig:thermo_classification}
\end{figure}

\textit{Type I: Null-null degenerate.}
Here $E_{\xi\xi}$ vanishes identically off-shell on a stationary Killing horizon and gives no information about $D_A\kappa$. GR, $f(R)$ gravity, and Lovelock theories are standard examples. This class naturally splits into two subclasses.

\textit{Type IA: Mixed projection sufficient.}
Here $E_{\xi A}$ is sufficient to establish the zeroth law non-perturbatively under the appropriate matter assumptions. GR and $f(R)$ gravity belong to this class. In the presence of matter, once the null-null flux is set to zero, the dominant energy condition implies the vanishing of the null-transverse flux as well. In $f(R)$ gravity one further requires $f'(R)\neq0$; the stronger condition $f'(R)>0$ follows from the positivity of the horizon entropy density.

\textit{Type IB: Mixed projection obstructed.} Here the mixed equation alone does not generically yield a non-perturbative proof. Lovelock gravity provides the canonical example, with the schematic form $\mathcal M^A{}_B D^B\kappa=0$. A non-perturbative argument therefore requires control over the invertibility of $\mathcal M^A{}_B$, which is not guaranteed in general. Otherwise, one is restricted to perturbative proofs on the Einstein branch~\cite{GhoshSarkar}.

\textit{Type II: Null-null nondegenerate.}
For this class, $E_{\xi\xi}$ does not vanish off-shell on a stationary Killing horizon. After imposing $T_{\xi\xi}=0$, it gives a nontrivial differential constraint on $\kappa$, schematically $E_{\xi\xi}=V(x)\,D^2(\kappa^2)+W^A(x)\,D_A(\kappa^2)+\cdots=0$, where $V(x)$ and $W^A(x)$ are smooth horizon fields obtained by restricting spacetime curvature quantities and their derivatives to $\cH$. The resulting equation need not be elliptic or close on $\kappa$ alone. When its principal part is elliptic, either automatically or with the aid of some additional geometric or physical conditions, compactness and regularity can nevertheless enforce $D_A\kappa=0$. The GQG result in Eq.~\eqref{eq:intro} is the simplest example, with $\{V=-\beta, \, W^A=0\}$.

A more instructive example is the cubic theory: $ {\cal L} = R+\lambda\, R R_{ab}R^{ab}$, for which the stationary-horizon null-null equation turns out to be
\[
E_{\xi\xi}=-\lambda\, D_A\!\left(RD^A\kappa^2\right).
\]
More details are provided in \ref{appB}. On each connected region where $R$ has a definite nonzero sign, this equation provides an elliptic constraint with $\{V=-\lambda\, R, \, W^A=-\lambda\, D_AR\}$ and enforces constancy of $\kappa$. If $R$ vanishes on an open region $\Omega_0$, the null-null equation is degenerate there. Remarkably, the mixed equation $E_{\xi A}$ then reduces to
\[
(1+\lambda\, R_{ab}R^{ab})\, D_A\kappa=0.
\]
Since the prefactor controls the local Wald entropy density on $\Omega_0$, $s_{\rm Wald}=(\sqrt \sigma/4G)\,(1+\lambda\, R_{ab}R^{ab})$~\cite{Wald:1993nt}, it is positive and enforces $D_A\kappa=0$ on $\Omega_0$. Then, a smooth matching across its boundary establishes the zeroth law globally. This provides a nontrivial Type-II realization in which the null-null equation controls the nondegenerate regions, while the mixed equation takes over precisely where it degenerates.

\textit{Type III: Generic.}
More general higher-curvature theories need not belong to either of the classes described above. For such theories, no general non-perturbative proof of the zeroth law is presently known. The most general result available is instead the perturbative proof of Ref.~\cite{Bhattacharyya:2022nqa}, which applies to diffeomorphism-invariant theories treated within an effective-field-theory expansion around solutions smoothly connected to GR. This completes the classification, which is summarized schematically in Fig.~\ref{fig:thermo_classification}.

\section{Discussion}

A central lesson of our analysis is that the zeroth law is enforced by the underlying field equations in a theory-dependent way. In GR, $f(R)$ gravity, and Lovelock theories, the null-null projection is degenerate and the standard argument is controlled by $E_{\xi A}$. In contrast, GQG behaves qualitatively differently: $E_{\xi\xi}$ itself becomes an intrinsic elliptic constraint, yielding an exact, branch-independent zeroth law.

The distinction becomes particularly instructive when the relevant horizon projection operator loses rank. In Lovelock gravity the mixed equation takes the schematic form $\mathcal M^{AB} D_B\kappa=0$, so a null eigenvalue of $\mathcal M^A_B$ marks a genuine degeneracy of the stationary horizon constraint: the mixed equation alone no longer controls the corresponding component of $D_A\kappa$. Importantly, this does not by itself imply a violation of the zeroth law; it only signals that additional horizon equations or physical input may become necessary. In the case of Lovelock gravity, the additional input is supplied by restricting the solution space to the Einstein branch of the theory. In contrast, for $(R+\lambda\, R R_{ab}R^{ab})$ theory discussed above, the mixed equation takes over where the null-null equation degenerates on $R=0$ patches.

There is also a suggestive thermodynamic interpretation of such degeneracies. For Lovelock gravity, the operator $\mathcal M^{AB}$ is proportional to the local metric response of the Jacobson-Myers entropy functional (see \ref{appC}),
\[
\mathcal M^{AB}\propto
\frac{1}{\sqrt \sigma}\frac{\delta S_{\rm JM}}{\delta \sigma_{AB}} .
\]
Thus, a null eigenvalue of the zeroth-law matrix is simultaneously a soft direction of the horizon entropy response. The connection is even more explicit in $(R+\lambda\, R R_{ab}R^{ab})$ theory, where the coefficient that restores the mixed zeroth-law equation on the degenerate region is precisely the coefficient multiplying the local Wald entropy density.

This loss of rank should be distinguished from the characteristic degeneracy of the full spacetime principal symbol. Killing horizons can already be characteristic in higher-curvature gravity, whereas the degeneracy discussed here concerns the stationary transverse constraint that enforces uniform surface gravity. Nevertheless, both phenomena emphasize that rank properties of curvature-dependent differential operators play a central role in horizon dynamics, suggesting an interesting connection with recent analyses of characteristic structure in higher-curvature theories~\cite{AliSuneeta}. It would be important to understand whether a degenerate zeroth-law branch can support a regular stationary solution with $D_A\kappa\neq0$, or whether the remaining horizon equations always restore equilibrium.

Thus, in modified gravity, a uniform BH temperature is not guaranteed by stationarity alone, but emerges from how the gravitational field equations constrain horizon structure. The zeroth law therefore becomes more than a statement of equilibrium: it provides a sharp probe of the dynamical and thermodynamic structure of gravity itself.

Beyond the theoretical analysis, an intriguing question is whether the zeroth law itself can eventually be confronted with observations, much as the second law has been tested. Any departure from uniform surface gravity would modify the near-horizon redshift structure and could, in principle, leave an imprint on radiation generated or scattered close to the horizon. This is particularly suggestive for proposed ``direct-wave'' or other near-horizon contributions to the post-merger signal, as well as for precision ringdown spectroscopy~\cite{Oshita:2025qmn}. Although a practical zeroth-law test remains to be developed, such observables may ultimately probe whether a BH horizon has settled into a state of genuine thermal equilibrium.



\section{Acknowledgments}
R.G. is supported by the Fulbright Nehru Postdoctoral Research Fellowship (Award No.3174/FNPDR/2025) from the United States-India
Educational Foundation. R.G. is supported by NSF Grants No.~AST-2606672, No.~PHY-2513337, No.~PHY-090003, and No.~PHY-20043, by John Templeton Foundation Grant No.~62840, by the Simons Foundation [MPS-SIP-00001698, E.B.], by the Simons Foundation International [SFI-MPS-BH-00012593-02], and by Italian Ministry of Foreign Affairs and International Cooperation Grant No.~PGR01167. S.S.’s research is supported by the Department of
Science and Technology, Government of India, under the
ANRF CRG Grant (No. CRG/2023/000934).

\appendix
\labelformat{section}{Appendix #1} 
\labelformat{subsection}{Appendix \thesection.#1}

\section{Generalized quadratic gravity} \label{appA}

For ${\cal L}_{\rm GQG}=R-2\Lambda+\alpha R^2+\beta R_{ab}R^{ab}
+\gamma {\cal L}_{\rm GB}$, all but the Ricci-squared part of the field equation vanishes off-shell when the
$\xi\xi$-projection is taken on $\cH$ and the horizon identities in Eq.~\eqref{EM:horizon} are used. The surviving Ricci-squared piece produces an Euler tensor
\begin{align}
E^{(R_{cd}R^{cd})}_{ab}={}&2R_{acbd}R^{cd}+\Box R_{ab}-\nabla_a\nabla_bR
\nonumber\\[-1mm]
&\qquad+\frac12g_{ab}(\Box R-R_{cd}R^{cd}),
\label{ERicci}
\end{align}
In the null-null projection, the terms proportional to $g_{ab}$ vanish, and so does the third term
$\xi^a\xi^b\nabla_a\nabla_bR=0$, due to $\mathcal{L}_\xi R=0$. Whereas the first and second terms in the field equation contribute non-trivially. 

To simplify the first term in Eq.~\eqref{ERicci}, we start by decomposing $R_{\xi c \xi d}$ in the horizon basis $\{\xi^a,\, N^a,\, e^a_A\}$:
\begin{equation*}
    R_{\xi c \xi d} = R_{\xi N \xi N}\, \xi_c\, \xi_d + 2\, (D^A \kappa)\, \xi_{(c}e_{d) A},
\end{equation*}
which then implies $2R_{\xi c \xi d} R^{c d} = -4\, (D \kappa)^2$, using the horizon identities in Eq.~\eqref{EM:horizon}. Whereas for the second term $\Box R_{ab} \equiv g^{cd} \nabla_c \nabla_d R_{ab}$, we use the decomposition $g^{ab}=- 2\xi^{(a}N^{b)} + e_A^ae_B^b\sigma^{AB}$ and simplify term-by-term using Eq.~\eqref{EM:horizon}, to obtain $\xi^a\xi^b\Box R_{ab}=-2\kappa(D^2 \kappa)+2\,(D \kappa)^2$. Hence, putting all the contributions together, we get
\begin{align*}
E^{\rm GQG}_{\xi\xi}=-2\,\beta\,\kappa\, (D^2\kappa)-2\, \beta\, (D\kappa)^2= -\beta\,D^2(\kappa^2),
\end{align*}
as quoted in the main text.

\section{A degenerate Type-II example}\label{appB}
We now focus on the cubic theory: ${\cal L}=R+\lambda R R_{ab}R^{ab}$ and write its field equation as
$E_{ab}=G_{ab}+\lambda E^{(3)}_{ab}=8\pi G T_{ab}$, where
\begin{align}
&E^{(3)}_{ab}=S\, G_{ab}+2\,R\, R_{ac}\,R_b^c+g_{ab}\,[\Box S + \nabla_c\nabla_d(R\, R^{cd})]\nonumber\\
&-\nabla_a\nabla_bS+\Box(R\,R_{ab})-\nabla_c\nabla_a(R\,R_b^c)-\nabla_c\nabla_b(R\,R_a^c).
\label{EM:E3}
\end{align}
with $S \equiv R_{ab}R^{ab}$. In the null-null projection, all $g_{ab}$ terms vanish and
$R_{\xi\xi}=0$. Moreover, since $\mathcal{L}_\xi S =0$, the $\nabla_a\nabla_bS$ term also drops.
Hence, we have
\begin{align}
E^{(3)}_{\xi\xi}=2\,R\, R_{\xi c}\,R_\xi^c
+\xi^a\,\xi^b\, [\Box(X_{ab})-2\,\nabla_c\nabla_a(X_b^c)],
\label{EM:nullsplit}
\end{align}
with $X_{ab} \equiv R\,R_{ab}$.

Now, we shall make use of the contracted Bianchi identity and the derivative-commutation relations, to obtain two useful identities
\begin{align*}
&\nabla_c X_b^c=R_b^c\,\nabla_cR+\frac{R}{2}\,\nabla_b R, \\
&\nabla_c\nabla_aX_b^c
=\nabla_a\nabla_cX_b^c+R_{da}X_b^d
-R^d{}_{bca}X_d^c .
\end{align*}
Resolving these derivatives
in the null frame $\{\xi^a,\,N^a,\,e^a_A\}$ and using ${\cal L}_\xi R={\cal L}_\xi
R_{ab}=0$ give the intermediate identity
\begin{align}
\xi^a\xi^b\!\left[\Box X_{ab}
-2\nabla_c\nabla_aX_b^c\right]
=2D_A(\kappa R R_\xi^A)
-2R R_{\xi c}R_\xi^c .
\label{EM:keycancel}
\end{align}
Substituting Eq.~\eqref{EM:keycancel} into Eq.~\eqref{EM:nullsplit} makes the
cancellation transparent:
\begin{align*}
E^{(3)}_{\xi\xi}=2\,D_A(\kappa\, R\, R_\xi^A)=-D_A\!\left(R\, D^A\kappa^2\right).
\label{EM:weighted}
\end{align*}
Here, we have used $R_\xi^A=-D^A\kappa$ and
$2\kappa D^A\kappa=D^A(\kappa^2)$.
Finally, since $G_{\xi\xi}=T_{\xi \xi}=0$ on $\cH$, the field equation reduces to $D_A(R\,D^A\kappa^2)=0$, assuming $\lambda \neq 0$. 

On every connected nodal domain $\Omega\subset\{R\neq0\}$ of the non-extremal smooth horizon, Ricci scalar has a fixed sign and an integration by parts of $\kappa^2\, D_A(R\,D^A\kappa^2)=0$ implies
\begin{equation}
\int_\Omega\sqrt \sigma\,R\,(D_A \kappa^2)^2 = 0,
\end{equation}
where the boundary term vanishes because $R=0$ on $\partial\Omega$.
Thus, $D_A\kappa^2=0$, and hence, $D_A\kappa$ must vanish on each such domain.

However, if $R=0$ on an open region $\Omega_0$, the null-null equation degenerates. And, we also have
\begin{equation*}
R=0,\quad D_AR=0,\quad {\cal L}_\xi R=0.
\end{equation*}
So, on $\Omega_0$, we can write $\nabla_aR=-\xi_a \nabla_N R$. The Hessian
components entering Eq.~\eqref{EM:E3} then reduce to
\begin{align}
\nabla_A\nabla_BR&=0,\qquad \nabla_\xi\nabla_AR=0,\nonumber\\
\nabla_\xi\nabla_NR&=\kappa\, \nabla_N R,\qquad
\Box R=-2\, \kappa\, \nabla_N R.
\label{EM:hessR}
\end{align}
We now project Eq.~\eqref{EM:E3} along $\xi^a e_A^b$. The explicit algebraic
terms proportional to $R$ vanish, while $SR_{ab}$ contributes $SR_{\xi A}$.
For the derivative sector, we expand before setting $R=0$:
\begin{align*}
\Box(RR_{\xi A})={}&(\Box R)R_{\xi A}
 +2(\nabla^cR)\nabla_cR_{\xi A},\\
\nabla_c\nabla_\xi(RR_A^c)={}&(\nabla_c\nabla_\xi R)R_A^c
 +(\nabla_\xi R)\nabla_cR_A^c\\
& +(\nabla_cR)\nabla_\xi R_A^c,\\
\nabla_c\nabla_A(RR_\xi^c)={}&(\nabla_c\nabla_A R)R_\xi^c
 +(\nabla_A R)\nabla_cR_\xi^c\\
& +(\nabla_cR)\nabla_A R_\xi^c .
\end{align*}
On $\Omega_0$, $\nabla_A R=\nabla_\xi R=0$ and
$\nabla_aR=-\xi_a \nabla_N R$. Using Eq.~\eqref{EM:hessR}, the surviving terms
proportional to $\nabla_N R$ and $\kappa \nabla_N R$ all cancel between these three derivative
structures. Therefore, no normal derivative of $R$ remains, and
\begin{equation*}
E^{(3)}_{\xi A}=S\, R_{\xi A}\quad \text{on}\, \, \Omega_0.
\label{EM:mixedcubic}
\end{equation*}
Since $G_{\xi A}=R_{\xi A}=-D_A\kappa$, the full mixed equation
$E_{\xi A}=0$ (for $T_{\xi A}=0$) becomes
\begin{equation}
\left(1+\lambda R_{ab}R^{ab}\right)D_A\kappa=0.
\label{EM:mixed}
\end{equation}
The prefactor $\left(1+\lambda R_{ab}R^{ab}\right)$ also appears in the local Wald entropy on
$\Omega_0$:
\begin{equation*}
s_{\rm Wald}=\frac{\sqrt \sigma}{4G}
\left(1+\lambda R_{ab}R^{ab}\right) .
\end{equation*}
Pointwise positivity of $s_{\rm Wald}$ makes Eq.~\eqref{EM:mixed}
nondegenerate and gives $D_A\kappa=0$ on $\Omega_0$. Smooth matching to the
nodal domains then establishes the zeroth law globally.

\section{Lovelock constraint and entropy response} \label{appC}

For Lovelock gravity, $E_{\xi\xi}$ vanishes off-shell on a stationary
Killing horizon $\cH$ and the mixed equation has the form ${\cal M}_A^B D_B\kappa=0.$ In particular, for Einstein-Gauss-Bonnet gravity, we have
${\cal M}_A^B=\delta_A^B-4\,\gamma\,{}^{(D-2)}G_A^B$, where $\gamma$ is the Gauss-Bonnet coupling constant and ${}^{(D-2)}G_A^B$ is the intrinsic Einstein tensor of the horizon cross section $\cS$.
The Jacobson-Myers entropy
\begin{equation*}
S_{\rm JM}=\frac{1}{4G}\int_{\mathcal S}\sqrt \sigma\,
(1+2\,\gamma\,{}^{(D-2)}R)
\end{equation*}
has a variation, up to an intrinsic total derivative,
\begin{equation*}
\delta S_{\rm JM}=\frac{1}{8G}\int_{\mathcal S}\sqrt \sigma\,
\left(\sigma^{AB}-4\,\gamma\,{}^{(D-2)}G^{AB}\right)\delta \sigma_{AB}\, .
\end{equation*}
Therefore, we get
\begin{equation*}
{\cal M}^{AB}=\frac{8G}{\sqrt \sigma}
\frac{\delta S_{\rm JM}}{\delta \sigma_{AB}} .
\end{equation*}
A zero eigenvalue of ${\cal M}_A^B$ is thus simultaneously a soft
direction of the first-order entropy response. It signals a degeneracy of
the stationary horizon constraint, not by itself a violation of the zeroth law or a degeneracy of the full spacetime principal symbol. A similar argument follows for higher order Lovelock theories as well.



\begin{thebibliography}{99}
\bibitem{Wald:1999vt}
R.~M.~Wald,
Living Rev. Rel. \textbf{4}, 6 (2001)
doi:10.12942/lrr-2001-6
[arXiv:gr-qc/9912119 [gr-qc]].

\bibitem{Carlip:2014pma}
S.~Carlip,
Int. J. Mod. Phys. D \textbf{23}, 1430023 (2014)
doi:10.1142/S0218271814300237
[arXiv:1410.1486 [gr-qc]].

\bibitem{Wall:2018ydq}
A.~C.~Wall,
[arXiv:1804.10610 [gr-qc]].

\bibitem{Sarkar:2019xfd}
S.~Sarkar,
Gen. Rel. Grav. \textbf{51}, no.5, 63 (2019)
doi:10.1007/s10714-019-2545-y
[arXiv:1905.04466 [hep-th]].


\bibitem{BardeenCarterHawking}
J.~M.~Bardeen, B.~Carter, and S.~W.~Hawking,
The four laws of black hole mechanics,
Commun.\ Math.\ Phys.\ \textbf{31}, 161 (1973).



\bibitem{Hawking:1974rv}
S.~W.~Hawking,
Nature \textbf{248}, 30-31 (1974)
doi:10.1038/248030a0


\bibitem{Hawking:1975vcx}
S.~W.~Hawking,
Commun. Math. Phys. \textbf{43}, 199-220 (1975)
[erratum: Commun. Math. Phys. \textbf{46}, 206 (1976)]
doi:10.1007/BF02345020









\bibitem{Yunes:2009ke}
N.~Yunes and F.~Pretorius,
Phys. Rev. D \textbf{80}, 122003 (2009)
doi:10.1103/PhysRevD.80.122003
[arXiv:0909.3328 [gr-qc]].

\bibitem{Agathos:2013upa}
M.~Agathos, W.~Del Pozzo, T.~G.~F.~Li, C.~Van Den Broeck, J.~Veitch and S.~Vitale,
Phys. Rev. D \textbf{89}, no.8, 082001 (2014)
doi:10.1103/PhysRevD.89.082001
[arXiv:1311.0420 [gr-qc]].


\bibitem{Berti:2015itd}
E.~Berti, E.~Barausse, V.~Cardoso, L.~Gualtieri, P.~Pani, U.~Sperhake, L.~C.~Stein, N.~Wex, K.~Yagi and T.~Baker, \textit{et al.}
Class. Quant. Grav. \textbf{32}, 243001 (2015)
doi:10.1088/0264-9381/32/24/243001
[arXiv:1501.07274 [gr-qc]].


\bibitem{LIGOScientific:2016lio}
B.~P.~Abbott \textit{et al.} [LIGO Scientific and Virgo],
Phys. Rev. Lett. \textbf{116}, no.22, 221101 (2016)
[erratum: Phys. Rev. Lett. \textbf{121}, no.12, 129902 (2018)]
doi:10.1103/PhysRevLett.116.221101
[arXiv:1602.03841 [gr-qc]].

\bibitem{Isi:2019aib}
M.~Isi, M.~Giesler, W.~M.~Farr, M.~A.~Scheel and S.~A.~Teukolsky,
Phys. Rev. Lett. \textbf{123}, no.11, 111102 (2019)
doi:10.1103/PhysRevLett.123.111102
[arXiv:1905.00869 [gr-qc]].


\bibitem{LIGOScientific:2021sio}
R.~Abbott \textit{et al.} [LIGO Scientific, VIRGO and KAGRA],
Phys. Rev. D \textbf{112}, no.8, 084080 (2025)
doi:10.1103/PhysRevD.112.084080
[arXiv:2112.06861 [gr-qc]].


\bibitem{LIGOScientific:2026qni}
A.~G.~Abac \textit{et al.} [LIGO Scientific, VIRGO and KAGRA],
[arXiv:2603.19019 [gr-qc]].


\bibitem{LIGOScientific:2026oim}
A.~G.~Abac \textit{et al.} [LIGO Scientific, VIRGO and KAGRA],
[arXiv:2607.19293 [gr-qc]].

\bibitem{Isi:2020tac}
M.~Isi, W.~M.~Farr, M.~Giesler, M.~A.~Scheel and S.~A.~Teukolsky,
Phys. Rev. Lett. \textbf{127}, no.1, 011103 (2021)
doi:10.1103/PhysRevLett.127.011103
[arXiv:2012.04486 [gr-qc]].


\bibitem{LIGOScientific:2025rid}
A.~G.~Abac \textit{et al.} [LIGO Scientific, Virgo and KAGRA],
Phys. Rev. Lett. \textbf{135}, no.11, 111403 (2025)
doi:10.1103/kw5g-d732
[arXiv:2509.08054 [gr-qc]].


\bibitem{Rincon-Ramirez:2026tbo}
M.~Rincon-Ramirez, N.~K.~Johnson-McDaniel, E.~Bianchi, I.~Gupta, V.~Prasad and B.~S.~Sathyaprakash,
Phys. Rev. Lett. \textbf{137}, no.2, 021406 (2026)
doi:10.1103/hvp6-ydbq
[arXiv:2601.22388 [gr-qc]].


\bibitem{Jacobson:1995uq}
T.~Jacobson, G.~Kang and R.~C.~Myers,
Phys. Rev. D \textbf{52}, 3518-3528 (1995)
doi:10.1103/PhysRevD.52.3518
[arXiv:gr-qc/9503020 [gr-qc]].


\bibitem{GhoshSarkar}
R.~Ghosh and S.~Sarkar,
Black hole zeroth law in higher curvature gravity,
Phys.\ Rev.\ D \textbf{102}, 101503(R) (2020).

\bibitem{Dey:2021rke}
S.~Dey, K.~Bhattacharya and B.~R.~Majhi,
Phys. Rev. D \textbf{104}, no.12, 124038 (2021)
doi:10.1103/PhysRevD.104.124038
[arXiv:2105.07787 [gr-qc]].


\bibitem{Sang:2021rla}
A.~Sang and J.~Jiang,
Phys. Rev. D \textbf{104}, no.8, 084092 (2021)
doi:10.1103/PhysRevD.104.084092
[arXiv:2110.00903 [gr-qc]].


\bibitem{Fang:2022nfa}
C.~Fang, L.~Xie, J.~Jiang and M.~Zhang,
Phys. Lett. B \textbf{831}, 137149 (2022)
doi:10.1016/j.physletb.2022.137149
[arXiv:2205.04266 [gr-qc]].


\bibitem{Bhattacharyya:2022nqa}
S.~Bhattacharyya, P.~Biswas, A.~Dinda and N.~Kundu,
JHEP \textbf{10}, 013 (2022)
doi:10.1007/JHEP10(2022)013
[arXiv:2205.01648 [hep-th]].


\bibitem{Davies:2024fut}
I.~Davies,
Phys. Rev. D \textbf{109}, no.8, 084051 (2024)
doi:10.1103/PhysRevD.109.084051
[arXiv:2401.13075 [gr-qc]].


\bibitem{Sang:2024tjd}
A.~Sang and P.~Zhao,
Phys. Lett. B \textbf{856}, 138868 (2024)
doi:10.1016/j.physletb.2024.138868


\bibitem{Lu:2015cqa}
H.~Lu, A.~Perkins, C.~N.~Pope and K.~S.~Stelle,
Phys. Rev. Lett. \textbf{114}, no.17, 171601 (2015)
doi:10.1103/PhysRevLett.114.171601
[arXiv:1502.01028 [hep-th]].


\bibitem{Lu:2015psa}
H.~L{\"u}, A.~Perkins, C.~N.~Pope and K.~S.~Stelle,
Phys. Rev. D \textbf{92}, no.12, 124019 (2015)
doi:10.1103/PhysRevD.92.124019
[arXiv:1508.00010 [hep-th]].


\bibitem{Pravda:2016fue}
V.~Pravda, A.~Pravdova, J.~Podolsky and R.~Svarc,
Phys. Rev. D \textbf{95}, no.8, 084025 (2017)
doi:10.1103/PhysRevD.95.084025
[arXiv:1606.02646 [gr-qc]].


\bibitem{Kokkotas:2017zwt}
K.~Kokkotas, R.~A.~Konoplya and A.~Zhidenko,
Phys. Rev. D \textbf{96}, 064007 (2017)
doi:10.1103/PhysRevD.96.064007
[arXiv:1705.09875 [gr-qc]].


\bibitem{Podolsky:2019gro}
J.~Podolsk{\'y}, R.~{\v{S}}varc, V.~Pravda and A.~Pravdova,
Phys. Rev. D \textbf{101}, no.2, 024027 (2020)
doi:10.1103/PhysRevD.101.024027
[arXiv:1907.00046 [gr-qc]].


\bibitem{Held:2022abx}
A.~Held and J.~Zhang,
Phys. Rev. D \textbf{107}, no.6, 064060 (2023)
doi:10.1103/PhysRevD.107.064060
[arXiv:2209.01867 [gr-qc]].

\bibitem{note}
The Schwarzschild metric nevertheless remains the unique Einstein-branch solution outside any spherically symmetric configuration of QG, see Ref.~\cite{Ghosh:2024tlk} for more details.


\bibitem{Ghosh:2024tlk}
R.~Ghosh, A.~K.~Mishra and A.~Chowdhury,
Phys. Lett. B \textbf{871}, 139974 (2025)
doi:10.1016/j.physletb.2025.139974
[arXiv:2411.09193 [gr-qc]].

\bibitem{Padmanabhan:2013xyr}
T.~Padmanabhan and D.~Kothawala,
Phys. Rept. \textbf{531}, 115-171 (2013)
doi:10.1016/j.physrep.2013.05.007
[arXiv:1302.2151 [gr-qc]].


\bibitem{SarkarBhattacharya}
S.~Sarkar and S.~Bhattacharya,
The issue of zeroth law for Killing horizons in Lanczos--Lovelock gravity,
Phys.\ Rev.\ D \textbf{87}, 044023 (2013).


\bibitem{Edelstein:2024jzu}
J.~D.~Edelstein, R.~Ghosh, A.~Laddha and S.~Sarkar,
[arXiv:2409.16935 [hep-th]].


\bibitem{Jost}
J.~Jost,
\textit{Riemannian Geometry and Geometric Analysis},
7th ed. (Springer, Cham, 2017).


\bibitem{Racz:1995nh}
I.~Racz and R.~M.~Wald,
Class. Quant. Grav. \textbf{13}, 539-553 (1996)
doi:10.1088/0264-9381/13/3/017
[arXiv:gr-qc/9507055 [gr-qc]].


\bibitem{Wald:1993nt}
R.~M.~Wald,
Phys. Rev. D \textbf{48}, no.8, R3427-R3431 (1993)
doi:10.1103/PhysRevD.48.R3427
[arXiv:gr-qc/9307038 [gr-qc]].


\bibitem{AliSuneeta}
M.~Ali and V.~Suneeta,
Causal structure of higher curvature gravity,
Phys.\ Rev.\ D \textbf{112}, 024063 (2025).

\bibitem{Oshita:2025qmn}
N.~Oshita, S.~Ma, Y.~Chen and H.~Yang,
[arXiv:2509.09165 [gr-qc]].

\end{thebibliography}
\end{document}